\documentclass[%
 reprint,
superscriptaddress,
 amsmath,amssymb,
 aps,
]{revtex4-1}

\usepackage{graphicx}
\usepackage{dcolumn}
\usepackage{bm}

\begin{document}

\preprint{APS/123-QED}

\title{A Simple Method for Generating Electromagnetic Oscillations}
\author{V.A. Buts}
\email{vbuts@kipt.kharkov.ua}
\affiliation{National Science Center Kharkov Institute of Physics and Technology,
1 Akademicheskaya Str.,Kharkov 61008, Ukraine}
\affiliation{V.N. Karazin Kharkov National University,
Svobody Sq. 4, Kharkov 61022, Ukraine}
\author{D.M. Vavriv}
\email{vavriv@rian.kharkov.ua}
\affiliation{Department of Microwave Electronics, Institute of Radio Astronomy of NAS of Ukraine, 4 Chervonopraporna Str., Kharkov 61002, Ukraine}
\author{O.G. Nechayev}
\affiliation{Department of Microwave Electronics, Institute of Radio Astronomy of NAS of Ukraine, 4 Chervonopraporna Str., Kharkov 61002, Ukraine}
\author{D.V. Tarasov}
\affiliation{National Science Center Kharkov Institute of Physics and Technology,
1 Akademicheskaya Str.,Kharkov 61008, Ukraine}

\date{\today}

\begin{abstract}
We propose a novel approach to the generation of electromagnetic oscillations by means of a low-frequency pumping of two coupled linear oscillators. A theory of such generation mechanism is proposed, and its feasibility is demonstrated by using coupled RLC oscillators. A comparison of the theoretical results and the experimental data is presented.
\begin{description}
\item[PACS numbers]
42.65.Yj, 05.45.Xt, 72.30.+q, 07.57.-c
\end{description}
\end{abstract}
\maketitle
\section{\label{sec:level1}INTRODUCTION\protect\\}

The generation of a high-frequency electromagnetic oscillation by means of a
conversion of a low-frequency oscillation is widely used in many electronics
applications. The frequency multiplication and the parametric frequency
transformation are well known examples of such approach. However, these and
other similar solutions are principally based on the usage of nonlinear
elements or media for the frequency conversion. This complicates practical
realization of the known frequency conversion schemes with increasing the
required output frequency.

In this paper, we demonstrate that the generation of electromagnetic oscillations is possible by using nonreciprocally coupled linear oscillators only, where the coupling coefficient is modulated in time at a frequency, which is significantly lower than the natural frequencies of the oscillators. The electromagnetic oscillations excited at the frequencies that are close to the natural frequencies. The usage of linear oscillators simplifies practical realizations of such generators.

It should be noted that the excitation of high-frequency oscillations by low-frequency pumping of coupled linear oscillators is unexpected and new result.  It was common opinion that only frequency that coincides with the frequency of an external forcing can arise in such system. However, as it is shown in the paper, the presence of a nonreciprocal coupling between the oscillators changes dramatically the system dynamics.

In the next section, we study dynamics of such system theoretically.  The mechanism of the excitation of high-frequency oscillations is described in terms of so called second-order resonances. Such resonances have been studied so far mainly from the point of view of various dynamical systems [1 - 7] stability.

Results of the corresponding experimental study are described in section III. Coupled RLC-oscillators made of lumped elements were used in the experiments. The natural frequencies of the oscillators and, consequently, the frequencies of the output oscillation are around 20 \emph{MHz}. A stable high-frequency excitation was observed starting from the modulation frequency of about 0.4 \emph{MHz}. The comparison of the theoretical and experimental data is presented in this section as well. Section IV contains discussions and conclusions.

\section{THEORY}

The dynamics of two coupled linear oscillators is described by the following
system of ordinary differential equations for normalized coordinates
$x_{1}$ and $x_{2}$
\begin{equation}
\label{eq1}
\frac{d^{2}x_{k} }{dt^{2}}+\delta_{k} \frac{dx_{k} }{dt}+\nu_{k}^{2} x_{k}
+\mu_{k} (t)x_{j} =0, \; \kappa =1,2,
\end{equation}
where $j=1$ if $k=2$, and $j=2$ if $k=1$, $\delta_{k}$ are the damping
coefficients, $\nu_{k}$ are the partial frequencies, $\mu_{k}(t)$ are
the coupling coefficients.

It is well known that in the case of the lossless oscillators ($\delta_{1}
=\delta_{2}=$0) and a constant coupling ($\mu_{1}(t)=\mu_{01}
=const$, $\mu_{2}(t)=\mu_{02}=const)$, the system (1) is
characterized by the following natural frequencies
\begin{equation}
\label{eq2}
\omega_{1,2}^{2} =[\nu_{1}^{2} +\nu_{2}^{2} \pm \sqrt {(\nu_{1}^{2} -\nu
_{2}^{2} )^{2}+4\mu_{10} \mu_{20} } ]/2
\end{equation}

For our study, it is important that the frequencies $\omega_{1}$ and
$\omega_{2}$ are different even for the case of identical oscillators:
$\nu_{1}=\nu_{2}$ and $\mu_{1}=\mu_{2}$.

In order to determine other properties of the system (1), we introduce a new
variable
\begin{equation}
\label{eq3}
z_{k} =x_{k} \exp (\delta_{k} t/2), \; k=1,2,
\end{equation}
and rewrite equations (1) in the form:

\begin{equation}
\label{eq4}
\frac{d^{2}z_{k} }{dt^{2}}+v^{2}(1+F)z_{k} =\mu_{k} (t)z_{j} ,
\end{equation}
where $F=\delta^{2} /(4\nu^{2})$. For simplicity, we suppose here that
$\nu_{1} =\nu_{2} \equiv v$ and $\delta_{1} =\delta_{2} \equiv
\delta $

The above equations can be further simplified by assuming a small coupling
between the oscillators. This allows applying the method of averaging [8].
According to this method, we introduce a complex amplitude
$B_{k} (t)$, which is related to $z_{k} (t)$ in the following way:
\begin{equation}
\label{eq5}
z_{k} =B_{k} (t)\exp [i\nu t+iFt/(2\nu )],
\end{equation}
where $i=\sqrt {-1} $.

The corresponding system of the averaged equations reads
\begin{equation}
\label{eq6}
\frac{dB_{k}}{dt}=\frac{i}{2\nu} \mu_{k} B_{j} , \; k=1,2,
\end{equation}
where, like in (1) and (3), $j=1$ if $k=2$, and $j=2$ if $k=1$.

By using the above equations it is easy to find that
\begin{gather}
\label{eq7}
\frac{d}{dt}\left( {\left| {B_{1} } \right|^{2}+\left| {B_{2} } \right|^{2}}
\right)=\\
\nonumber =2\left[ {\mu_{1} (t)-\mu_{2} (t)} \right]Im\left[ {B_{1}
B_{2}^{\ast } \exp (i\Phi_{12} } \right].
\end{gather}
From (7) we note that if the coupling between the oscillators is reciprocal
($\mu_{1}(t)=\mu_{2}(t))$, than
\begin{equation}
\label{eq8}
\frac{d}{dt}(|\emph{B}_{1}|^{2}+|\emph{B}_{2}|^{2})=const
\end{equation}
Since $|\emph{B}_{1}|^{2}+|\emph{B}_{2}|^{2})=const$ is
proportional to the energy stored in the oscillators, we can conclude from
(7) and (8) that a reciprocal pumping prevents an energy increase in the
coupled oscillators. It is also possible to prove that such conclusion is
valid for an arbitrary number of nonidentical oscillators.

So, we can expect an excitation of high-frequency oscillations only if $\mu
_{1}(t)\neq\mu_{2}(t)$. For simplicity, we consider a nonreciprocal coupling
between the oscillators when one of the coefficients $(\mu_{1})$
changes with time and another one is constant:
\begin{gather}
\label{eq9}
    \mu_{1}(t)=\mu_{10}[1+m\cos (\Omega t)\\
    \nonumber \mu_{2}(t)=\mu_{20}
\end{gather}

Here $m$ is the modulation coefficients, $\Omega$ is the frequency of the
modulation.

Now we rewrite the system of equations (6) as a second order differential
equation:
\begin{equation}
\label{eq10}
\frac{d^{2}B_{2}}{dt^{2}}+\frac{\mu_{10} \mu_{20} }{4v^{2}}[1+m\cos(\Omega t)]B_{2} =0
\end{equation}
So we got a Mathieu-type equation, which properties are well known. In
particular, provided that the modulation coefficient $m$ is small and the following
resonant condition holds
\begin{equation}
\label{eq11}
\Omega \approx \sqrt{\mu_{10} \mu_{20}}/\nu ,
\end{equation}
the amplitude $B_{2}$ increases as
\begin{equation}
\label{eq12}
B_{2} \propto \exp \frac{m\sqrt {\mu_{10} \mu_{20} } }{8\nu }
\end{equation}
Taking into account the relation of the amplitude $B_{2} $with the initial
coordinates $x_{1}$ and $x_{2}$ (see (3) and (5)), we find that the excitation
of high-frequency oscillations takes place if the following threshold
condition is fulfilled:
\begin{equation}
\label{eq13}
\frac{m\sqrt {\mu_{10} \mu_{20} } }{4\nu }>\delta .
\end{equation}
The increment $(\alpha)$ of the oscillation build-up is
\begin{equation}
\nonumber
\alpha =\frac{m\sqrt {\mu_{10} \mu_{20} } }{8\nu }-\frac{\delta}{2}.
\end{equation}
In order to better appreciate the condition (11), we note that according to
(2) the difference of the natural frequencies ($\Delta \omega \equiv \omega
_{1} -\omega_{2} )$ for the considered case is
\[
\Delta \omega =\frac{\sqrt {\mu_{10} \mu_{20} } }{v},
\]
and instead of (11) we have
\begin{equation}
\label{eq14}
\Omega \approx \Delta \omega .
\end{equation}

Thus, for the excitation of the high-frequency oscillations the frequency
of the modulation should be approximately equal to the difference between
the natural frequencies of the oscillators. This is the condition for the
existence of the secondary resonance in the system (1).

Theoretical criteria (14) for the beginning of the generation coincide rather
well with the one following from computer simulations and experiments. Let us,
at first, consider results of numerical simulations of (1). The bifurcation
diagram of this system on the parameter plane of the frequency of modulation
versus the coefficient of the modulation is presented on Fig. 1 for different
values of the coupling coefficients $(\mu_{01} =\mu_{02} \equiv
\mu_{0})$. Note that in the simulation, we put $\nu =1$. The corresponding
values of the difference frequency $\Delta \omega$ are indicated by arrows.
It is clear that the minimum threshold for the generation in terms of the
modulation coefficient is realized if the condition (14) is fulfilled.
The threshold is decreasing with increasing the permanent coupling between
the oscillators, as it also follows from (13).

\begin{figure}[h!]
\includegraphics[width=\columnwidth]{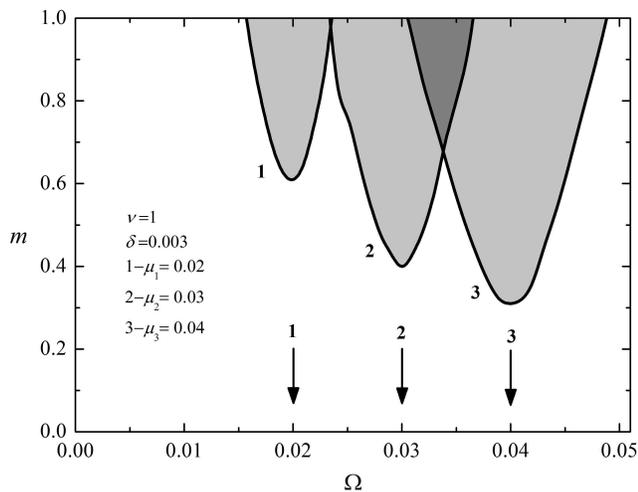}
\caption{\label{Fig1} Regions of parameters wherein the excitation of high-frequency  oscillations take place (gray).}
\end{figure}

The time evolution of the oscillations in the system is illustrated in Fig.
2. for the parameters of Fig 1, $\Omega$=0.04 and $m=$0.6. The intensity of the
excited oscillations increases exponentially. There is no saturation of the intensity
since(1) does not account for a particular nonlinearity that can be responsible for
the intensity saturation. Dissipative nonlinearity, anharmonizm of the oscillators,
or limiting power of the pumping oscillator can lead to a saturation of the intensity
 in a practical implementation of such generator.

\begin{figure}[h!]
\includegraphics[width=\columnwidth]{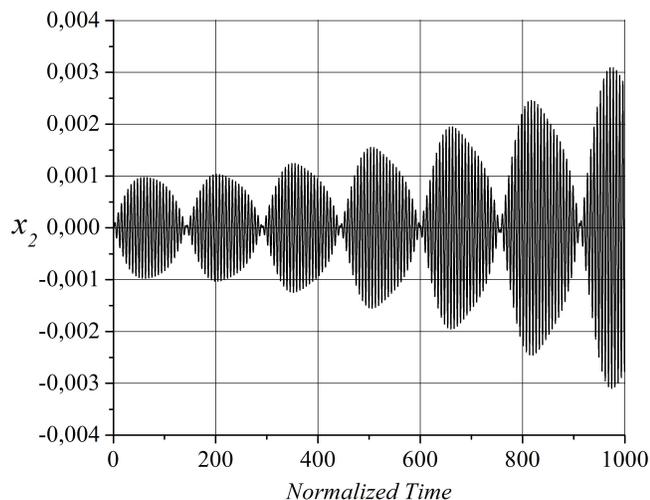}
\caption{\label{Fig2} Build-up of the oscillation.}
\end{figure}

From Fig. 2, we note that the excited oscillation is modulated. This
modulation take places at the frequency $\Delta \omega $. This is the
manifestation of the fact that a two-frequency oscillation is simultaneously
excited with the frequencies close to the natural frequencies $\omega_{1} $
and $\omega_{2} $. Since $\left| {\omega_{1} -\omega_{2} }
\right|<<\omega_{1} ,\omega_{2} $, the superposition of the oscillations
results in the observed modulation (beating).

\section{EXPERIMENT}

For the first demonstration of the feasibility of the proposed generation
mechanism we used two coupled RLC-oscillators as shown in Fig. 3. The partial
frequencies of the oscillators were equal to 19.935 \textit{MHz}, and the Q-factors of
the unloaded first and second oscillators were 200 and 220, respectively. A
coupling between the first (L1, C1) and the second (L2,C2)
\begin{figure}
\includegraphics[width=\columnwidth]{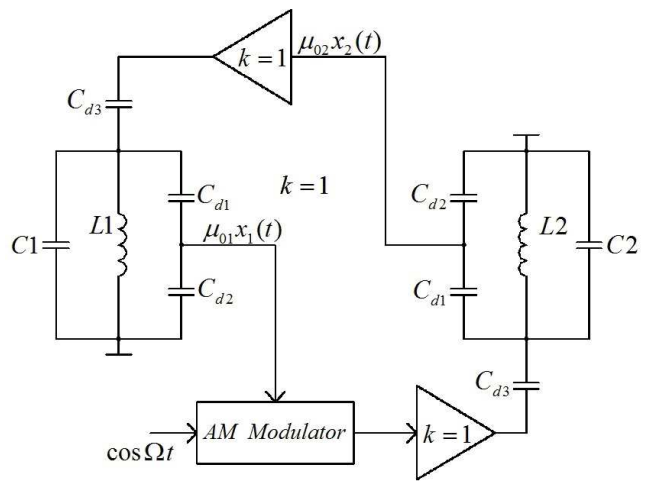}
\caption{\label{Fig3} Circuit diagram of the generator.}
\end{figure}
oscillators is provided by using a capacitor coupler (Cd1, Cd2, Cd3), a
buffer circuit and a modulator. The buffer circuit, implemented using IC AD8009,
acts as an isolator providing the transfer coefficient is equal to 0 dB and -20
dB in the forward and reverse direction, respectively. The modulator, based
on the FET BF998, is introduced to provide the modulation of the coupling
coefficient. The modulator is driven by an external low-frequency
oscillator.

The coupling from the second to the first oscillator is provided in the same
manner, but the modulator is not introduced here.

We measured the natural frequencies of the coupled oscillators when the
modulation was not applied. These frequencies are 19.656 MHz and 20.214 MHz.
For symmetrical and constant coupling ($\mu_{01} =\mu_{02} \equiv \mu_{0}
)$  and for the indicated value of $\nu$ it gives $\mu_{0} /v^{2}$=0.29. Such
value of the coupling coefficients follows also from the characteristics of
the capacitor dividers used in the experimental set up in Fig. 3.

An example of a comparison of the experimental and theoretical bifurcation
diagrams is given in Fig. 4. In the experiment, the minimum modulation
coefficient for the generation onset is reached at the modulation frequency
of about 0.58\textit{MHz}, which is rather close to the measured value of $\Delta \omega $.

A typical waveform and the spectrum of the output oscillation is shown on Fig. 5 and
Fig. 6, respectively. There are two intensive spectral components in the spectrum at
the frequencies that are rather close to the natural frequencies indicated above.
The superposition of such harmonic components results in the modulation of the output
oscillation envelope as it is seen from Fig. 6. The output power in our experiment is
limited by the power of the low-frequency oscillator used for the pumping.

From our point of view, the experimental results completely confirm the results following
 from the presented theoretical study.
\begin{figure}
\includegraphics[width=\columnwidth]{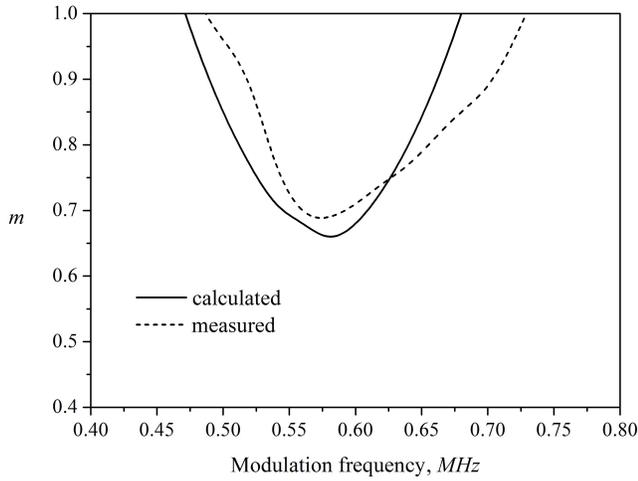}
\caption{\label{Fig4} Calculated and measured boundaries of the arias, where the excitation of the oscillation takes place.}
\end{figure}
\begin{figure}
\includegraphics[width=\columnwidth]{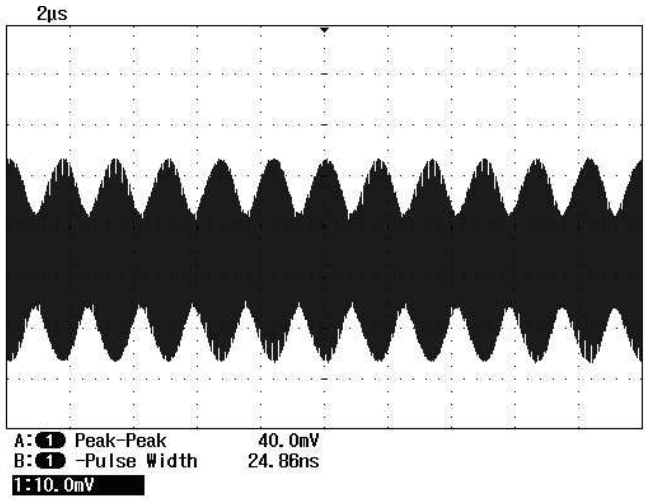}
\caption{\label{Fig5} Waveform of the excited oscillation.}
\end{figure}
\begin{figure}
\includegraphics[width=\columnwidth]{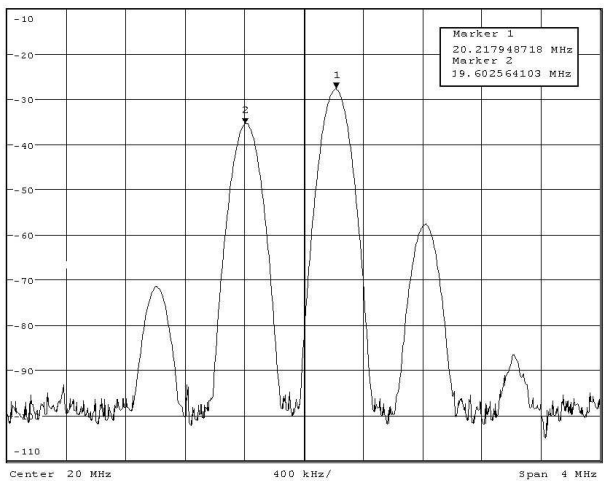}
\caption{\label{Fig6} Spectrum of the excited oscillation.}
\end{figure}

\section{DISCUSSION AND CONCLUSION}

The described mechanism of the generation of high-frequency oscillations can
be explained by using an energy diagram shown in Fig. 7. There are two
energy levels $\hslash \omega_{1}$ and $\hslash \omega_{2}$ that are
determined
\begin{figure}
\includegraphics[width=\columnwidth]{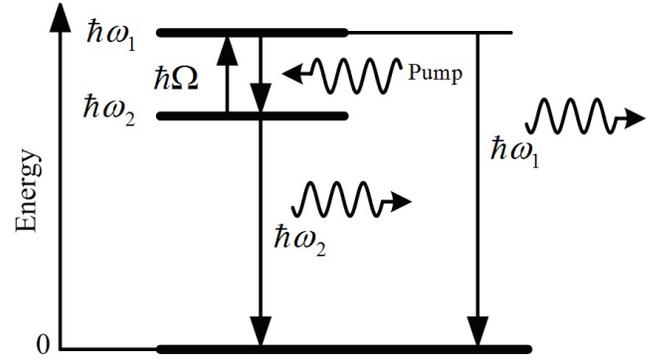}
\caption{\label{Fig7} Diagram of the generation  in the pumped coupled oscillators.}
\end{figure}
by the natural frequencies of the coupled oscillators. The width of the
levels are respectively $\hslash \omega_{1} /Q_{1} $ and $\hslash \omega
_{2} /Q_{2} $, where $Q_{1}$ and $Q_{2}$ are quality factors of the
oscillators. When pumping with the energy $\hslash \Omega $ is applied, the
transition
\begin{equation}
\label{eq15}
\hslash \omega_{2}+\hslash \Omega \rightarrow \hslash \omega_{1}
\end{equation}
occurs leading to increasing the population of the higher-lying level. Next,
since there is a coupling between the higher-lying and the lower-lying
level, the transition
\begin{equation}
\label{eq16}
\hslash \omega_{1}\rightarrow \hslash \omega_{2}
\end{equation}
also takes place. This process is accompanied by the radiation on the
frequency $\omega_{1} -\omega_{2}$. But the most important is that due to
(\ref{eq1}), a feedback is realized in the system. This feedback is the condition
to realize a self-excitation of high-frequency oscillations.

Next, since the coupling between the oscillators is not reciprocal, there is
always a population inversion of the levels necessary for the
development of the processes (15) and (16). The population of the levels is
saturated due to a nonlinear effect.

When a load is connected, the system oscillates at the
frequencies $\omega_{1} $, $\omega_{2} $, and their combinations, what is
also seen from the spectrum of output oscillation in Fig. 6.

We have experimentally demonstrated the suggested excitation mechanism
by using conventional lamped electronic components. It allows us to excite
oscillations at the frequencies around 20 \emph{MHz} by the external pumping
at the frequency of 0.58 \emph{MHz}. Such realization is convenient from the
point of view of reliable and detailed comparison of the experimental data and the theoretical results.
\nopagebreak
The proposed method for generating electromagnetic oscillations is very simple
since it does not require sophisticated nonlinear elements or media for its realization.
Therefore, this mechanism is especially interesting for the development of generators
for much higher frequencies.  For example, the availability of high-Q \emph{THz} resonators [9]
and variety of Ka-band oscillators available for the pumping and modulators create a solid background
for the development of \emph{THz} generators based on the proposed approach.
\nocite{*}

\bibliography{ButsVavriv}

\providecommand{\noopsort}[1]{}\providecommand{\singleletter}[1]{#1}%
\begin{thebibliography}{9}%
\makeatletter
\providecommand \@ifxundefined [1]{%
 \@ifx{#1\undefined}
}%
\providecommand \@ifnum [1]{%
 \ifnum #1\expandafter \@firstoftwo
 \else \expandafter \@secondoftwo
 \fi
}%
\providecommand \@ifx [1]{%
 \ifx #1\expandafter \@firstoftwo
 \else \expandafter \@secondoftwo
 \fi
}%
\providecommand \natexlab [1]{#1}%
\providecommand \enquote  [1]{``#1''}%
\providecommand \bibnamefont  [1]{#1}%
\providecommand \bibfnamefont [1]{#1}%
\providecommand \citenamefont [1]{#1}%
\providecommand \href@noop [0]{\@secondoftwo}%
\providecommand \href [0]{\begingroup \@sanitize@url \@href}%
\providecommand \@href[1]{\@@startlink{#1}\@@href}%
\providecommand \@@href[1]{\endgroup#1\@@endlink}%
\providecommand \@sanitize@url [0]{\catcode `\\12\catcode `\$12\catcode
  `\&12\catcode `\#12\catcode `\^12\catcode `\_12\catcode `\%12\relax}%
\providecommand \@@startlink[1]{}%
\providecommand \@@endlink[0]{}%
\providecommand \url  [0]{\begingroup\@sanitize@url \@url }%
\providecommand \@url [1]{\endgroup\@href {#1}{\urlprefix }}%
\providecommand \urlprefix  [0]{URL }%
\providecommand \Eprint [0]{\href }%
\providecommand \doibase [0]{http://dx.doi.org/}%
\providecommand \selectlanguage [0]{\@gobble}%
\providecommand \bibinfo  [0]{\@secondoftwo}%
\providecommand \bibfield  [0]{\@secondoftwo}%
\providecommand \translation [1]{[#1]}%
\providecommand \BibitemOpen [0]{}%
\providecommand \bibitemStop [0]{}%
\providecommand \bibitemNoStop [0]{.\EOS\space}%
\providecommand \EOS [0]{\spacefactor3000\relax}%
\providecommand \BibitemShut  [1]{\csname bibitem#1\endcsname}%
\let\auto@bib@innerbib\@empty
\bibitem [{\citenamefont {Lichtenberg}\ and\ \citenamefont
  {Lieberman}(1983)}]{1}%
  \BibitemOpen
  \bibfield  {author} {\bibinfo {author} {\bibfnamefont {A.~J.}\ \bibnamefont
  {Lichtenberg}}\ and\ \bibinfo {author} {\bibfnamefont {M.~A.}\ \bibnamefont
  {Lieberman}},\ }\href@noop {} {\emph {\bibinfo {title} {Regular and
  stochastic motion}}}\ (\bibinfo  {publisher} {Springer-Verlag New York
  Heidelberg Berlin},\ \bibinfo {year} {1983})\ p.\ \bibinfo {pages}
  {499}\BibitemShut {NoStop}%
\bibitem [{\citenamefont {Nayfeh}\ \emph {et~al.}(1994)\citenamefont {Nayfeh},
  \citenamefont {Nayfeh},\ and\ \citenamefont {Balachandran}}]{2}%
  \BibitemOpen
  \bibfield  {author} {\bibinfo {author} {\bibfnamefont {A.~H.}\ \bibnamefont
  {Nayfeh}}, \bibinfo {author} {\bibfnamefont {S.~A.}\ \bibnamefont {Nayfeh}},
  \ and\ \bibinfo {author} {\bibfnamefont {B.}~\bibnamefont {Balachandran}},\
  }\enquote {\bibinfo {title} {Transfer of energy from high-frequency to low
  frequency modes},}\ in\ \href@noop {} {\emph {\bibinfo {booktitle}
  {Nonlinearity and Chaos in Engineering Dynamics}}},\ \bibinfo {editor}
  {edited by\ \bibinfo {editor} {\bibfnamefont {J.~M.~T.}\ \bibnamefont
  {Thompson}}\ and\ \bibinfo {editor} {\bibfnamefont {S.~R.}\ \bibnamefont
  {Bishop}}}\ (\bibinfo  {publisher} {John Wiley and Sons Ltd},\ \bibinfo
  {year} {1994})\ pp.\ \bibinfo {pages} {39--58}\BibitemShut {NoStop}%
\bibitem [{\citenamefont {Oksasoglu}\ and\ \citenamefont {Vavriv}(1994)}]{3}%
  \BibitemOpen
  \bibfield  {author} {\bibinfo {author} {\bibfnamefont {A.}~\bibnamefont
  {Oksasoglu}}\ and\ \bibinfo {author} {\bibfnamefont {D.~M.}\ \bibnamefont
  {Vavriv}},\ }\href@noop {} {\bibfield  {journal} {\bibinfo  {journal} {IEEE
  Trans. Circuits and Systems.}\ }\textbf {\bibinfo {volume} {41}},\ \bibinfo
  {pages} {669} (\bibinfo {year} {1994})}\BibitemShut {NoStop}%
\bibitem [{\citenamefont {Vavriv}\ \emph {et~al.}(1996)\citenamefont {Vavriv},
  \citenamefont {Ryabov}, \citenamefont {Sharapov},\ and\ \citenamefont
  {Ito}}]{4}%
  \BibitemOpen
  \bibfield  {author} {\bibinfo {author} {\bibfnamefont {D.~M.}\ \bibnamefont
  {Vavriv}}, \bibinfo {author} {\bibfnamefont {V.~B.}\ \bibnamefont {Ryabov}},
  \bibinfo {author} {\bibfnamefont {S.~A.}\ \bibnamefont {Sharapov}}, \ and\
  \bibinfo {author} {\bibfnamefont {H.~M.}\ \bibnamefont {Ito}},\ }\href@noop
  {} {\bibfield  {journal} {\bibinfo  {journal} {Physical Review E}\ }\textbf
  {\bibinfo {volume} {53}},\ \bibinfo {pages} {103} (\bibinfo {year}
  {1996})}\BibitemShut {NoStop}%
\bibitem [{\citenamefont {Shygimaga}\ \emph {et~al.}(1998)\citenamefont
  {Shygimaga}, \citenamefont {Vavriv},\ and\ \citenamefont {Vinogradov}}]{5}%
  \BibitemOpen
  \bibfield  {author} {\bibinfo {author} {\bibfnamefont {D.~V.}\ \bibnamefont
  {Shygimaga}}, \bibinfo {author} {\bibfnamefont {D.~M.}\ \bibnamefont
  {Vavriv}}, \ and\ \bibinfo {author} {\bibfnamefont {V.}~\bibnamefont
  {Vinogradov}},\ }\href@noop {} {\bibfield  {journal} {\bibinfo  {journal}
  {IEEE Trans. Circuits and Systems}\ }\textbf {\bibinfo {volume} {45}},\
  \bibinfo {pages} {1255} (\bibinfo {year} {1998})}\BibitemShut {NoStop}%
\bibitem [{\citenamefont {Buts}(2001)}]{6}%
  \BibitemOpen
  \bibfield  {author} {\bibinfo {author} {\bibfnamefont {V.~A.}\ \bibnamefont
  {Buts}},\ }\href@noop {} {\bibfield  {journal} {\bibinfo  {journal} {Problems
  of Atomic Science and Technology. Special issue dedicated to the 90-th
  birthday anniversary of A.I. Akhiezer.}\ }\textbf {\bibinfo {volume} {6}},\
  \bibinfo {pages} {329} (\bibinfo {year} {2001})}\BibitemShut {NoStop}%
\bibitem [{\citenamefont {Buts}(2004)}]{7}%
  \BibitemOpen
  \bibfield  {author} {\bibinfo {author} {\bibfnamefont {V.~A.}\ \bibnamefont
  {Buts}},\ }\href@noop {} {\bibfield  {journal} {\bibinfo  {journal}
  {Electromagnetic waves and electron systems (In Russian)}\ }\textbf {\bibinfo
  {volume} {9}},\ \bibinfo {pages} {59} (\bibinfo {year} {2004})}\BibitemShut
  {NoStop}%
\bibitem [{\citenamefont {Bogoliubov}\ and\ \citenamefont
  {Mitropolski}(1961)}]{8}%
  \BibitemOpen
  \bibfield  {author} {\bibinfo {author} {\bibfnamefont {N.~N.}\ \bibnamefont
  {Bogoliubov}}\ and\ \bibinfo {author} {\bibfnamefont {Y.~A.}\ \bibnamefont
  {Mitropolski}},\ }\href@noop {} {\emph {\bibinfo {title} {Asymptotic Methods
  in the Theory of Nonlinear Oscillations}}}\ (\bibinfo  {publisher} {Gordon
  and Breach, New York},\ \bibinfo {year} {1961})\BibitemShut {NoStop}%
\bibitem [{\citenamefont {Yang}\ \emph {et~al.}(2012)\citenamefont {Yang},
  \citenamefont {Katz}, \citenamefont {Willis}, \citenamefont {Weber},
  \citenamefont {Knezevic}, \citenamefont {Hagness},\ and\ \citenamefont
  {Booske}}]{9}%
  \BibitemOpen
  \bibfield  {author} {\bibinfo {author} {\bibfnamefont {B.~B.}\ \bibnamefont
  {Yang}}, \bibinfo {author} {\bibfnamefont {S.~L.}\ \bibnamefont {Katz}},
  \bibinfo {author} {\bibfnamefont {K.~J.}\ \bibnamefont {Willis}}, \bibinfo
  {author} {\bibfnamefont {M.~J.}\ \bibnamefont {Weber}}, \bibinfo {author}
  {\bibfnamefont {I.}~\bibnamefont {Knezevic}}, \bibinfo {author}
  {\bibfnamefont {S.~C.}\ \bibnamefont {Hagness}}, \ and\ \bibinfo {author}
  {\bibfnamefont {J.~H.}\ \bibnamefont {Booske}},\ }\href@noop {} {\bibfield
  {journal} {\bibinfo  {journal} {IEEE Trans. THz Science and Technology}\
  }\textbf {\bibinfo {volume} {2}},\ \bibinfo {pages} {449} (\bibinfo {year}
  {2012})}\BibitemShut {NoStop}%
\end{thebibliography}%

\end{document}